# Uncover the Dynamic Community Structure of Instant Delivery Network

Chengbo Zhang[1], Yonglin Li[1], Zuopeng Xiao[1]
[1]Harbin Institute of Technology
Email: {zhangcb0027; liyonglin2000; tacxzp}@foxmail.com

## Abstract

The rise of instant delivery services has reshaped urban spatial structures through the interaction between suppliers and consumers. However, limited research has explored the spatiotemporal dynamics of delivery network structures. This study constructs an time dependent multi-layer instant delivery network in case city Beijing using a large-scale dataset from Eleme, organized into 500m grid units. A dynamic community detection method identifies evolving community structures over time. Results reveal 309 dynamic communities, with an average size of 13.78 km². Communities form in the morning, expand, stabilize, then contract and disappear by night. Key factors influencing stability include building area and residential population, while online retail and service facilities contribute to instability. These findings offer insights into the spatial structure of instant delivery networks and the factors driving their dynamics, with practical implications for optimizing platform strategies, resource allocation, and urban transportation planning.

## 1. Introduction

Human activities such as population movement, freight transport, and information flows create complex spatial networks that are essential for understanding urban systems. These networks provide insights into urban spatial organization and its dynamics (Cao et al., 2023; Dong et al., 2024; Jia et al., 2021). The emergence of on-demand instant delivery services has introduced a new form of spatial interaction between suppliers and consumers, blending online-to-offline (O2O) service models. Instant delivery has revolutionized residents' access to daily services, influencing both spatial and temporal dimensions of (Bissell, 2020; Kong et al., 2024; Zhang et al., 2020). As such, instant delivery has become a key component of urban logistics, reshaping urban mobility and service accessibility.

Despite the critical role of instant delivery networks, two significant research gaps remain unexplored. **First,** the spatial structure of these networks is unclear. The flow of goods and riders in response to rapid on-demand services and regulated by platform policy, creates self-organizing urban clusters or communities (M. Wang et al., 2023; Yang et al., 2024, Kong et al., 2024; Z. Wang & He, 2021). These communities, marked by dense internal supply-demand connections, impact urban systems including delivery capacity and service accessibility (Li et al., 2024). Understanding the spatial patterns of instant delivery communities is essential for understanding urban space in digital era and efficient resource allocation. **Second,** instant delivery networks are inherently dynamic (Wen et al., 2022), with fluctuating consumer demand and platform strategies that cause these communities to continuously evolve. Figure 1 summarizes the spatial logic of this dynamic community structure. Static spatial analysis methods fail to capture the temporal shifts in these networks, which are essential for understanding variations in delivery demand and operational needs (Bovet et al., 2022; Rossetti & Cazabet, 2018). A dynamic network approach, particularly using dynamic community detection, can reveal the spatiotemporal evolution of these communities, identifying stable and fluctuating regions over time (Chen et al., 2024; Jia et al., 2022; Zhao et al., 2023). Such insights are valuable for optimizing platform strategies, adjusting resource allocation, and informing adaptive urban transportation policies to better align with real-time service demands.

To address the gaps above, we propose a dynamic community detection framework to analyze instant delivery networks. We construct the network using 500m × 500m grid units and apply a cross-layer dynamic community detection method combining the MutuRank and Leiden algorithms. This approach allows us to examine the evolving community structures and analyze how spatial factors influence their dynamics. Our study enhance the understanding of urban mobility system through the lens of instant delivery. Besides, it provides important practical insights for optimizing platform operations and urban planning, with implications for improving efficient and dynamic delivery regionalization.

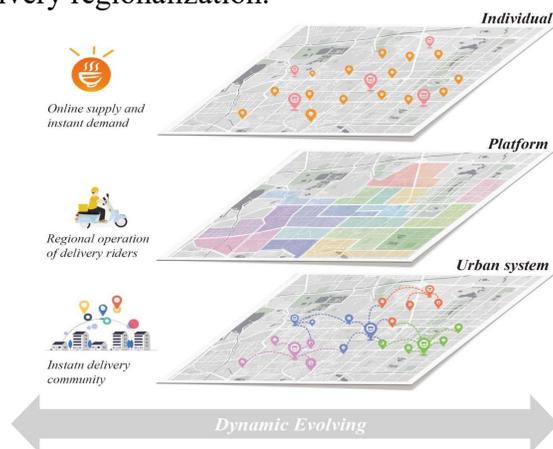

Figure 1. The spatial logic of instant delivery community structure

## 2. Methodology
### 2.1 Data
The dataset consists of 278,726 delivery orders sourced from Ele.me, a leading on-demand platform, covering February 2020 in Beijing, China. Despite the COVID-19 pandemic, daily activities remained consistent, ensuring the data's representativeness for typical delivery patterns. We aggregate the order flow into 500m × 500m grid cells, to create the instant delivery origin-destination (OD) network. Figure 2 displays the constructed network in the research area, Beijing.

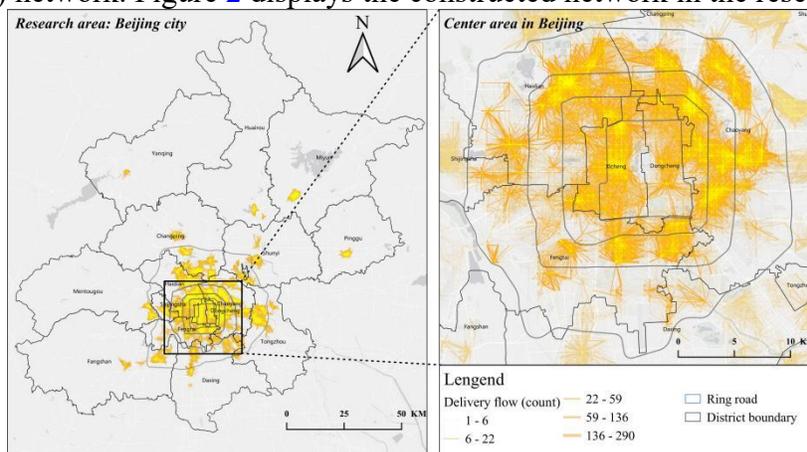

Figure 2. Research area and instant delivery OD flow aggregated in grids

## 2.2 Dynamic community detection and analysis

Referring to Jia et al. (2022), we propose a three-step community detection and analysis frameworks. Figure 3 illustrates the work flow. The first step is cross-time multi-layer networks construction. The delivery data is organized into 18 one-hour time slices, covering the period from 6:00 am to 12:00 midnight. Each delivery order flow is represented as an OD pair, where intra-layer edges connect orders within the same time slice, and inter-layer edges link orders across different time layers. This structure enables the analysis of both spatial proximity and temporal interactions in the network.

The second step is community detection on multi-layer networks. We use the random walk-based MutuRank algorithm to calculate the weight of each time layer, accounting for the influence of different time steps. The Leiden algorithm is then applied to the time-weighted multi-layer network to detect dynamic communities—well-connected groups of grid units across the entire network. These communities are identified within each time layer, allowing us to observe their temporal variation.

The final step is spatiotemporal evolution analysis. We analyze the life cycle of dynamic delivery communities using Jaccard similarity to quantify the overlap between communities across time layers. By tracking the temporal changes in community membership, we identify stable and fluctuating units throughout the day. Additionally, we explore how these variations relate to specific land use patterns, using explainable machine learning methods that combine XGBoost classification with SHAP for interpreting the relative contribution of factors.

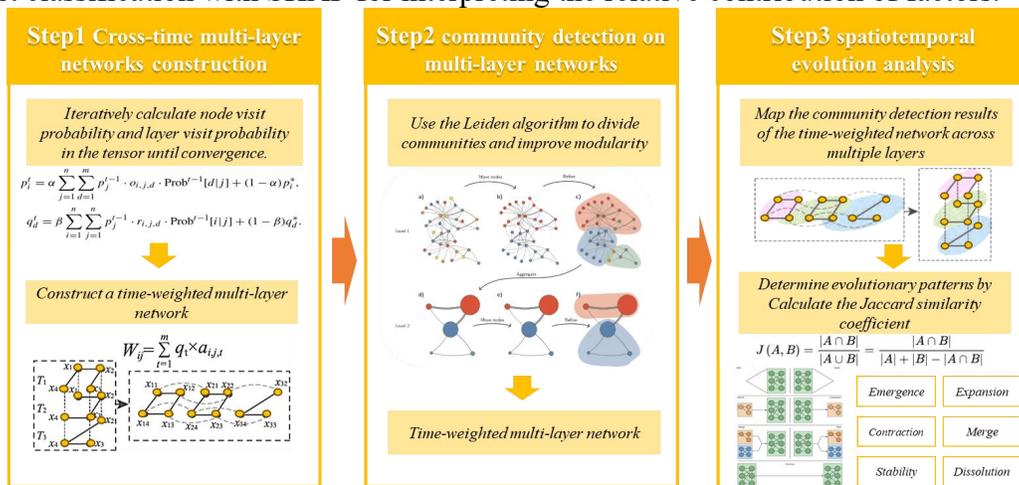

Figure 3. Work flow of dynamic community detection and analysis

## 3. Results

### 3.1 Dynamic instant delivery communities and its evolving patterns

We identified 309 dynamic instant delivery communities through time-layered analysis. Figure 4 shows snapshots of these communities across 18 time periods. Spatially, they exhibit a compact-to-sparse distribution, with dense communities in the city center and more dispersed ones in suburban areas like Huairou and Miyun districts. Notably, delivery communities extend beyond administrative boundaries, reflecting the fluid, human-centered nature of urban spatial organization. Figure 5 shows the distribution of the size and the number of dynamic communities across different time period. Specifically, the number of communities remains stable between 9:00 AM and 10:00 PM, while their size follows an "increase-decrease" pattern. The peak occurs between 11:00 and 12:00, with an average area of 13.78 km².

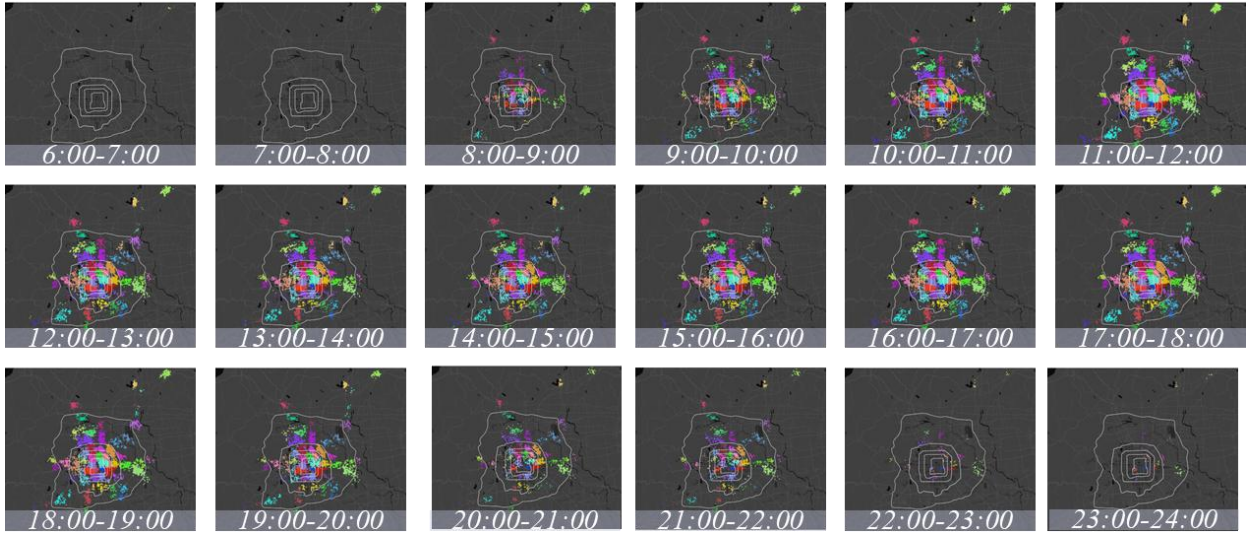

Figure 4. Distribution of dynamic instant delivery communities in each time period

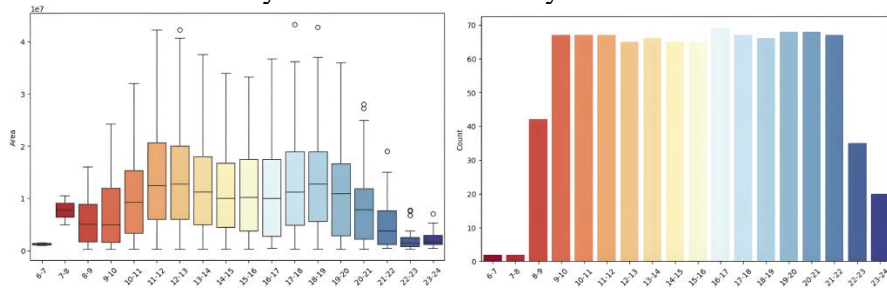

Figure 5. Distribution of the size (left) and the number (right) of dynamic communities across different time period

As Figure 6 shows, the life course of instant delivery communities follows five stages: birth, expansion, stabilization, contraction, and dissolution. Long-duration zones (13-14 hours) typically form in the morning, expand, stabilize, and dissolve by midnight, while short-duration zones (3-5 hours) form and dissolve quickly, often around noon or evening.

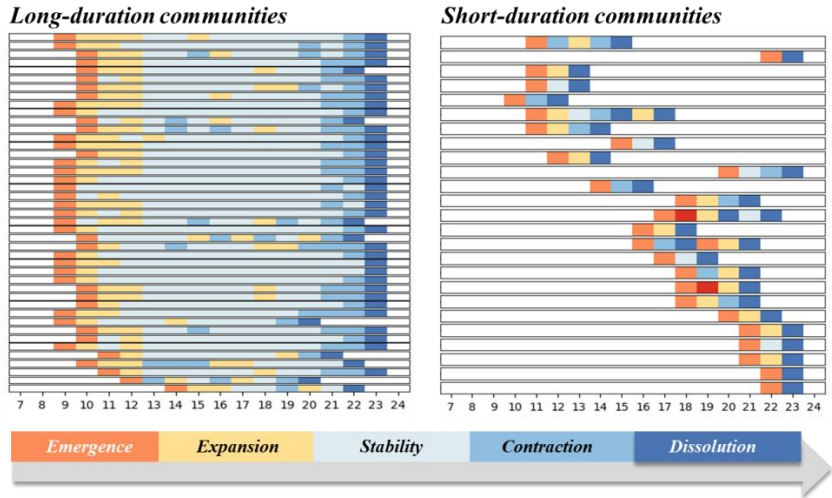

Figure 6. Life courses of dynamic instant delivery communities

## 3.2 The driven factors of variation of dynamic units

Figure 7 shows the distribution of variable and stable units, along with SHAP-based identification of driving factors. Observations indicate that stable units form specific spatial clusters, while variable units are distributed along certain boundaries. The relative importance analysis reveals significant contributions from building area and population, both positively correlated, suggesting that population aggregation leads to instability in service units due to population activity rhythms. Conversely, the number of service facilities, including online retail outlets, plays a stabilizing role, indicating that commercial clusters act as anchors for service units to some extent.

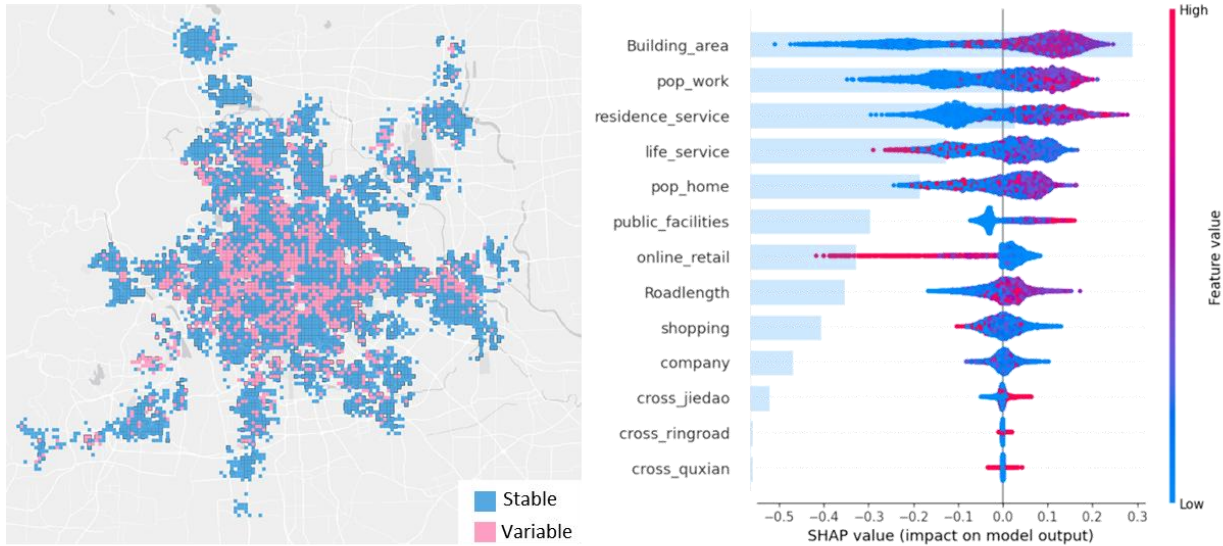

Figure 7. Identification of units variance (left) and the relative contribution of land use and facility factors on the variance (right).

## 4. Discussion

Based on the empirical findings, we discuss the influence of dynamic community structure on understanding and facilitating urban space, planning and platform strategies. Figure 8 displays these implications.

    **Dynamic Urban Structure through the lens of on-demand Service.** Instant delivery create flexible urban structures that transcend administrative boundaries, driven by demand patterns rather than fixed zones. This reflects "rescaling" and "re-territorialization" in the process platform urbanism (Caprotti & Liu, 2022; Talamini et al., 2022) as urban structure adapt dynamically to consumer needs, forming human-centered regions that shift in response to real-time activity.

    **Urban Function Shaped Community Evolution.** Our study reveals specific urban functions could shape the stability of delivery zones. Stable zones, often with abundant commercial facilities, exhibit consistent demand, while fluctuating zones emerge in areas with high mobility and variable demand. This highlights the influence of land use on delivery patterns, with stable areas meeting routine needs and fluctuating zones responding to more transient demands.

    **Sub-Regions for Capacity Optimization.** Identifying dynamic community structure sub-regions aids in platform optimization. This dynamic partitioning strategy, echoes the ideas of similar work in the field of operations research (Ouyang et al., 2022, 2023), and synergizes to help more effective capacity allocation and management.

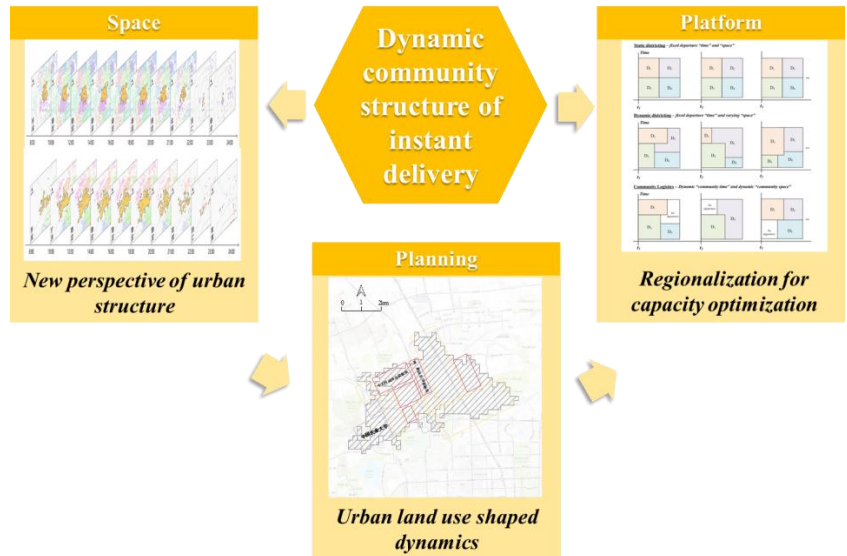

Figure 8. The influence of dynamic community structure on space, urban planning, and platform.